\documentclass[11pt]{article}%
\usepackage{amssymb, amsfonts, amsmath}
\usepackage{graphicx}
\usepackage{float}
\usepackage{geometry}%
\usepackage{mathtools}
\usepackage{nicefrac}
\usepackage{tabularx, booktabs, multirow}
\usepackage{tablefootnote}
\allowdisplaybreaks
\usepackage{epstopdf}
\usepackage{hyperref}
\usepackage{multicol}
\usepackage[natbib, style=apa]{biblatex}
\usepackage[dvipsnames]{xcolor}
\usepackage[ruled, vlined]{algorithm2e}  
\usepackage{subcaption}

\usepackage{pifont} 

\providecommand{\U}[1]{\protect\rule{.1in}{.1in}}
\newtheorem{theorem}{Theorem}[section]

\newtheorem{definition}{Definition}[section]

\newtheorem{lemma}{Lemma}[section]

\newtheorem{remark}{Remark}[section]

\numberwithin{equation}{section}

\numberwithin{equation}{section}

\newcommand{\be}{\begin{equation}}
\newcommand{\ee}{\end{equation}}

\newcommand{\bq}{\begin{eqnarray}}
\newcommand{\eq}{\end{eqnarray}}

\begin{document}

\title{Exploratory Mean-Variance with Jumps: An Equilibrium Approach}
\author{Yuling Max Chen \thanks{Department of Statistics and Actuarial Science, University of
Waterloo, Waterloo ON N2L 3G1, Canada (yuling.chen@uwaterloo.ca)}
\and Bin Li\thanks{Department of Statistics and Actuarial Science, University of
Waterloo, Waterloo ON N2L 3G1, Canada (bin.li@uwaterloo.ca)}
\and David Saunders \thanks{Department of Statistics and Actuarial Science, University of
Waterloo, Waterloo ON N2L 3G1, Canada (dsaunder@uwaterloo.ca).}}
\date{{\small \today}}
\maketitle

Revisiting the continuous-time Mean-Variance (MV) Portfolio Optimization problem, we model the market dynamics with a jump-diffusion process and apply Reinforcement Learning (RL) techniques to facilitate informed exploration within the control space. We recognize the time-inconsistency of the MV problem and adopt the time-inconsistent control (TIC) approach to analytically solve for an exploratory equilibrium investment policy, which is a Gaussian distribution centered on the equilibrium control of the classical MV problem. Our approach accounts for time-inconsistent preferences and actions, and our equilibrium policy is the best option an investor can take at any given time during the investment period. 
Moreover, we leverage the martingale properties of the equilibrium policy, design a RL model, and propose an Actor-Critic RL algorithm. 
All of our RL model parameters converge to the corresponding true values in a simulation study. Our numerical study on 24 years of real market data shows that the proposed RL model is profitable in 13 out of 14 tests, demonstrating its practical applicability in real world investment.

\vspace{-0.5cm}

\newpage

\section{Introduction} \label{sec:introduction}
The Mean-Variance (MV) portfolio optimization problem, first introduced by \citet{markowitz1952portfolio}, has been 
the subject of extensive research. 
In the continuous-time setting, an MV investor aims to optimize her financial outcome by pursuing the highest possible portfolio return with the lowest risk, measured by portfolio volatility, at the end of the investment horizon. 
\citet{jorion1992portfolio} investigated the practical challenges of implementing the MV investment policy in 
real-world markets. \citet{zhou2000continuous} studied the Lagrangian dual of the MV problem as a stochastic linear-quadratic optimal control problem. 
Later on, \citet{chiu2006asset} solved for the MV efficient frontier, and \citet{xie2008continuous} derived the optimal investment policy in an incomplete market setting; see \citet{zhang2018portfolio, kalayci2019comprehensive} for a broader overview of the the past studies of the MV problem. 

In this paper, we take the stochastic control approach and extend the classical MV problem in two directions. In the first direction, we employ a Reinforcement Learning (RL) facilitated {\it exploratory formulation} of the MV problem. 
In the recent stochastic control literature, there is an emerging trend of applying RL techniques to classical stochastic control problems \citep{wang2020continuous, wang2020reinforcement, jiang2022reinforcement, jia2022policy, PE, jia2023q, dai2023learning, wu2024reinforcement, chen2025EMVRS}.
The fundamental technique is to replace the optimal solution to the classical stochastic control problem with a probability distribution (centered around the original optimal solution), in light of the Stochastic RL Algorithm by \citet{gullapalli1990stochastic}.
Such a distribution-valued control is often called an {\it exploratory control} or a {\it policy} in the RL context. In this way, the exploratory control is capable of informed exploration within the control space, while maintaining greedy exploitation towards optimality. 
We point out here that the major motivation for such an exploratory formulation is the lack of knowledge of the controlled dynamics. Taking portfolio optimization as an example, the controlled dynamics is the portfolio value process that depends on the market dynamics. However, the precise law of the market evolution is unknown, and even a simple diffusion model requires knowledge of the drift and diffusion terms. 

\citet{wang2020reinforcement} first introduced this ``exploratory extension" to their controlled dynamics and solved for the optimal exploratory control of a linear-quadratic problem. 
They also proved the asymptotic equivalence between the classical solution and the exploratory solution.
Later, \citet{PE, jia2022policy, jia2023q} leveraged the martingale properties of the optimal solutions to the exploratory-extended stochastic control problems and developed effective learning algorithms for RL policy evaluation and policy updates.
\citet{wang2020continuous} studied the {\it Exploratory Mean-Variance} (EMV) problem, which solves for a distribution-valued investment policy centered on the classical MV solution.
Since the EMV problem uses the Lagrangian dual of the classical MV problem, it is a time-consistent control problem and the proposed investment policy in \citet{wang2020continuous} is only optimal at the beginning of the investment horizon. 

In contrast, \citet{dai2023learning} investigated the EMV problem as a time-inconsistent control (TIC) problem, as the time-inconsistency arises from the variance of terminal portfolio values, a quadratic function of the expected terminal portfolio value. Taking the TIC approach in \citep{bjork2021time}, \citet{dai2023learning} solved the Extended Hamiltonian-Jacobi-Bellman (EHJB) equations for an equilibrium policy, which is a probability distribution centered around the classical TIC solution provided in \citep{bjork2021time}. 
While \citet{dai2023learning} incorporated an exogenous and hidden state to model the unobservable market regimes or economic conditions, \citet{wu2024reinforcement} and \citet{chen2025EMVRS} modeled the market regimes more directly with a time-homogeneous Markov chain and solved the EMV problem with regime-switching.
Moreover, \citet{chen2025EMVRS} also pointed out the significance of model parameter convergence as an evaluation criterion for RL models. 

Our second extension to the classical MV problem is the incorporation of a jump component in the controlled dynamics, which is used to account for sudden shocks in the market. 
Jump-diffusion models have been widely applied to optimal stopping \citep{pham1997optimal, mordecki2002optimal, bayraktar2008optimizing}, option evaluation \citep{amin1993jump, xu2009jump, andersen2000jump, carr2007numerical, clift2008numerical, toivanen2008numerical}, and hedging problems \citep{park1993optimal, he2006calibration, mina2015approximate}. 
The MV problem with jump-diffusion controlled dynamics is solved in \citet{oksendal2005applied} in the classical stochastic control formulation. More recently, \citet{gao2024reinforcement} introduced the exploratory formulation of stochastic control problems with jump-diffusion controlled dynamics. However, both of them considered the Lagrangian dual of the MV problem, hence following the time-consistent control approach. 

Differing from \citet{gao2024reinforcement} and \citet{oksendal2005applied}, this paper tackles the EMV problem for a jump-diffusion by considering it as a TIC problem and solving for an equilibrium solution, an approach adopted by \citet{dai2023learning} and \citet{bjork2021time}. We establish the exploratory formulation of the MV problem with jump-diffusion market dynamics, which we call the {\it Exploratory Mean-Variance with Jumps (EMVJ) problem}.
We derive the EHJB equations and solve for an equilibrium investment policy in explicit form. Our solution aligns with \citet{bjork2021time} if we remove the exploration and the jumps in the dynamics. Following \citet{PE}, we then utilize the martingale properties of our analytical solutions and develop an Actor-Critic RL algorithm. Through iterative interactions with the market, our algorithm is able to gradually understand the market behavior and learn the equilibrium investment policy, without the knowledge of the underlying market dynamics. 
Our numerical results demonstrates the effectiveness of our RL algorithm on simulated data, as well as the practical applicability of our RL model on real market data. 

We make two main contributions to the literature. First, we fill a gap in the literature by solving the EMV problem with jump-diffusion dynamics via the TIC approach. We note that \citet{gao2024reinforcement}, as well as many other existing works \citep{wang2020continuous, gao2024reinforcement, wu2024reinforcement, chen2025EMVRS}, solved the Lagrangian dual of the MV problem. Their investment strategies are optimal at the beginning of an investment horizon subject to a prespecified investment target, disregarding the time-inconsistent preferences and actions an investor can have. Our equilibrium investment policy is a game-theoretical result, supposing an investor competes with her future incarnations throughout the investment horizon. Therefore, our equilibrium investment policy is the best action an investor can take, at any given time during the investment horizon. 
Second, we fully leverage our analytical solutions in the design of a RL model, by parameterizing the RL model with market parameters that have practical meanings. Our choice of parameterization not only enables us to check whether the parameters have converged to the true values in our simulation study, but also significantly simplifies the RL model\footnote{As described in Section \ref{sec:algorithm}, our RL model has only 3 parameters. } without sacrificing its effectiveness. 

\bigskip

The remainder of this paper is structured as follows. We review the necessary background and preliminaries in Section \ref{sec:preliminary}. In Section~\ref{sec:methodology}, 
we introduce the EMVJ problem and present the verification theorem and analytical solutions. We propose a RL algorithm in Section \ref{sec:algorithm} and present our numerical results in Section \ref{sec:numerical}. Finally, we conclude in Section \ref{sec:conclusion}.

\section{Problem Formulation and Preliminaries} \label{sec:preliminary}
We adopt the notation from \citep{oksendal2005applied}. 
Consider a finite decision horizon $T$ and a filtered probability space $(\Omega, \mathcal{F}, \{\mathcal{F}_t\}_{t\in  [0,T]}, \mathbb{P})$.
Let $L := \{L_t\}_{t\in[0,T]}$ be a L\'evy process and $\Delta L_s := L_s - L_{s-}$ be the jump of $L$ at time $s$. Denote by $\boldsymbol{B}_0$ the collection of Borel sets of $\mathbb{R}$ whose closure does not contain 0. Then, for $B \in \boldsymbol{B}_0$, define the Poisson random measure of $L$ as 
\begin{equation}
    N(t,B) = \sum_{s: 0 \leq s \leq t} 1_B(\Delta L_s), 
\end{equation}
and the L\'evy measure of $L$ as 
\begin{equation}
    \nu(B) = \mathbb{E}[N(1,B)].
\end{equation} 
Intuitively, $N(t,B)$ represents the number of jumps of sizes in a Borel set $B$ up to time $t$, whereas $\nu(B)$ measures the expected number of jumps of sizes in $B$ per unit time. For any $B \in \boldsymbol{B}_0, \{N(t,B): t \in [0,T]\}$ is a Poisson process with intensity $\nu(B)$. 

Define the compensated Poisson random measure as
\begin{equation}
    \Tilde{N}(dt,dz) = N(dt,dz) - \nu(dz) dt.
\end{equation}
We know from \citep{oksendal2005applied} that for any $B \in \boldsymbol{B}_0, \{\Tilde{N}(t,B): t \geq 0\}$ is a martingale. The L\'evy measure is $\nu(dz) = \zeta_J F_J(dz)$,

where $\zeta_J > 0$ is the arrival rate of Poisson jumps per unit time and $F_J(z)$ is the probability distribution of jump size $z$.
Here, we require that the L\'evy measure should satisfy 
\begin{equation} \label{eq:conditions-of-Levy-measure}
    \int_\mathbb{R} \min(1, z^2) \nu(dz) < \infty
    \quad \text{ and } 
    \int_{\mathbb{R}} e^{2z} \nu(dz) < \infty.
\end{equation}


For simplicity, we consider investing in a market with one risky asset (stock) and one risk-free asset. The risk-free rate is a constant $r \geq 0$. 
The stock price is driven by 
\begin{equation}
    \frac{dS_t}{S_{t-}} = \mu dt + \sigma dW_t + \int_\mathbb{R} (e^z-1) \Tilde{N}(dt,dz), 
\end{equation}
where $\mu \in \mathbb{R}, \sigma > 0$ are constant market parameters and $\{W_t\}_{t\in[0,T]}$ is a one-dimensional Brownian Motion on the filtered probability space. 
Under the second condition of (\ref{eq:conditions-of-Levy-measure}), we have 
that $\mathbb{E}[S_t]$ and $\mathbb{E}[S_t^2]$ are finite for any $t \in[0,T]$, according to \citep{cont2003financial}. 

An investor controls the money (in dollar value) that she invests in the stock, reallocating the portfolio in response to the behavior of the market. 
The discounted self-financing portfolio value $X := \{X_t\}_{t\in[0,T]}$ follows the process
\begin{equation}
    dX_t^u = u_t (\mu-r) dt + u_t \sigma dW_t + \int_\mathbb{R} u_t (e^z - 1) \Tilde{N}(dt,dz), 
\end{equation}
where $u := \{u_t := u(t,X_t^u)\}_{t\in[0,T]}$ denotes the control, i.e., money invested in the stock. 
 
\begin{remark}
Note that in reality the {\it market parameters}, $(\mu, \sigma)$, and the {\it jump parameters}, including $\zeta_J$ and the parameters of $F_J(z)$ (the jump size distribution), are unknown. So, investing under a control $u$, as a deterministic function of time $t$ and wealth $x$ that depends on the market parameters and the jump parameters, is impractical. While one possible solution is to estimate the unknown parameters from the market data, for example using Maximum Likelihood Estimation, this method often suffers from estimation error as the observed market data is just one realization of the market dynamics; see \citep{luenberger2013investment} for the {\it mean-blur problem} as a typical example. 
Alternatively, we can adopt the idea of Stochastic Reinforcement Learning \citep{gullapalli1990stochastic} that employs a random investor exploring for the optimal investment policy, within the control space. In this way, we avoid estimation error as there is no parameter estimation involved. Instead, the parameters are learned while the investor interacts with the dynamic market. In recent stochastic control literature, such an investor randomization is achieved by extending the classical control $u$ to a probability distribution valued policy, $\pi$, often centered at $u$; see \citep{wang2020continuous, wang2020reinforcement, jiang2022reinforcement, chen2025EMVRS, gao2024reinforcement, jia2022policy, jia2023q} for examples of such extensions. 
\end{remark}

Furthermore, we recognize the difficulty of investing under a control $u$ in practice, as the {\it market parameters}, $(\mu, \sigma)$, and the {\it jump parameters}, including $\zeta_J$ and the parameters of $F_J(z)$ (the jump size distribution), are unknown.
Henceforth, we consider an {\it exploratory control}, $\pi := \{\pi_{t, X_t^\pi}(u) := \pi(u; t, X_t^\pi)\}_{t\in[0,T]}$, that enables exploration within the control space. 
Here, $\pi:(t,x) \in [0,T] \times \mathbb{R} \mapsto \pi_{t,x} \in \mathcal{P}(\mathbb{R})$ is a stochastic feedback policy, which maps a time and portfolio value pair $(t,x)$ to a probability density function on the control space $\mathbb{R}$. 
For each $t \in [0,T]$, let $u_t^\pi := u(t,X_t^\pi)$ be the investment control sampled from $\pi_{t,X_t^\pi}(u)$. Then, a sampled portfolio value process is given by
\begin{equation} \label{eq:sampled-dX-pi}
    dX_t^\pi = u_t^\pi (\mu-r) dt + u_t^\pi \sigma dW_t + \int_\mathbb{R} u_t^\pi (e^z - 1) \Tilde{N}(dt,dz).
\end{equation}
To account for the stochasticity of the randomly sampled control, we consider an enlargement of the filtered probability space. Following \citep{jia2022policy, jia2023q}, we suppose that the probability space is rich enough to support mutually independent copies of a random variable, $\{U_t, t\in[0,T]\}$, each uniformly distributed on $[0,1]$. Denote by $(\Omega, \mathcal{F}, \{\mathcal{G}_t\}_{t\in[0,T]}, \Bar{\mathbb{P}})$ the new filtered probability space. 
Here, $\mathcal{G}_s = \mathcal{F} \vee \sigma(U_t, 0 \leq t \leq s)$ is the $\sigma$-algebra generated by $\mathcal{F}_s$ and the uniform random variables up to time $s$, and $\Bar{\mathbb{P}}$ is a product extension from $\mathbb{P}$, which is now defined on the extended $\sigma$-algebra $\mathcal{G}_T$. 
Following the arguments in \citep{gao2024reinforcement}, we also introduce an extended Poisson random measure, $N'(dt,dz,dv)$, on the product space $[0,T]\times \mathbb{R} \times [0,1]$, where $v$ is a realization of a Uniform random vector $V \sim \mbox{UNIF}([0,1])$. The intensity measure of $N'$ is given by $dt \, \nu(dz) dv$ and its compensator is $\Tilde{N'}(dt,dz,dv) := N'(dt,dz,dv) - dt \nu(dz) dv$.  
Furthermore, given a distribution $d$, denote $G^d: [0,1] \to \mathbb{R}$ by the quantile function of $d$. Then, for some $(t,x)$, the policy distribution is $\pi_{t,x}$ and a sampled action $u$ from policy can be realized by sampling $v \sim \mbox{UNIF}([0,1])$ and applying the quantile function, i.e., $u = G^{\pi_{t,x}}(v)$. 

Therefore, as we randomly sample trajectories of a control $u^\pi$ from a policy $\pi$ and take the average of the portfolio values across trajectories, the sampled portfolio value process in \ref{eq:sampled-dX-pi} follows the same distribution as
\begin{equation} \label{eq:dX-pi}
    dX_t^\pi = \int_{\mathbb{R}} u (\mu-r) \pi_{t, X_t^\pi}(u) du dt + \sqrt{\int_{\mathbb{R}} u^2 \sigma^2 \pi_{t, X_t^\pi}(u)du} dW_t + \int_{\mathbb{R} \times [0,1]} G^{\pi_{t,x}}(v) (e^z - 1) \Tilde{N'}(dt,dz, dv), 
\end{equation}
which we refer to as the {\it exploratory portfolio value process}.

\begin{definition}[Admissible Policy]
\label{def:admissible-policy} We say that $\pi :=\{\pi _{t,x}(\cdot
)\}_{t,x\in \lbrack 0,T]\times 
\mathbb{R}
_{+}}$ is an admissible policy if the following conditions are satisfied:

\begin{enumerate}
    \item[(1)] For any $(t,x)\in \lbrack 0,T]\times \mathbb{R}$, $\pi _{t,x}(u)\in \mathcal{P}(\mathbb{R})$;
    \item[(2)] 
    $\int_\mathbb{R} u^2 \pi_{t,x}(u) du \leq C_1(1+x^2)$, for any $(t,x) \in [0,T] \times \mathbb{R}$; 
    \item[(3)] $\left| \int_{\mathbb{R}} u(\pi_{t,x}(u) - \pi_{t,y}(u)) du \right|^2 + \left|\sqrt{\int_\mathbb{R} u^2 \pi_{t,x}(u) du } - \sqrt{\int_\mathbb{R} u^2 \pi_{t,y}(u) du }\right|^2 
    + \int_{[0,1]} |G^{\pi_{t,x}}(v) - G^{\pi_{t,y}}(v)|^2 dv \\
    \leq C_2|x-y|^2$
    , for any $t \in [0,T], x,y \in \mathbb{R}$;
    \item[(4)] $\int_{\mathbb{R}}\pi _{t,x}(u)|\log \pi
    _{t,x}(u)|du < C_{3}(1+|x|^{2})$, for any $(t,x)\in [0,T]\times \mathbb{R}$;
    
\end{enumerate}
Here, $C_{1},C_{2},C_{3}$ are constants that are independent of $t$ and $x$. The set of all admissible policies is denoted by $\mathcal{A}.$
\end{definition}

We conclude this section with the following lemma, which connects the definition of an admissible policy to the solution of the SDE (\ref{eq:dX-pi}). The proof is given in Appendix \ref{apdx:proof-of-lemma-admissible-policy-to-SDE-solution}
\begin{lemma} \label{lm: admissible-policy-to-SDE-solution}
Following an admissible policy $\pi \in \mathcal{A}$, the exploratory portfolio value process (\ref{eq:dX-pi}) guarantees a unique c\'adl\'ag adapted solution $X^\pi$ such that 
\begin{equation*}
    \mathbb{E}\left[\sup_{t\in[0,T]} (X_t^\pi)^2 \right] < \infty. 
\end{equation*}
\end{lemma}


\section{Exploratory Mean-Variance with Jumps Problem} \label{sec:methodology}
The mean-variance portfolio optimization problem aims to find an investment policy that achieves the highest return and the lowest portfolio volatility. 
Since the investment control $u$ is randomly sampled from the exploratory control $\pi$, we consider entropy regularization to encourage exploration within the control space. Denote by $H(\pi) := - \int_\mathbb{R} \pi(u) \log \pi(u) du$ the entropy of a distribution-valued exploratory policy $\pi$. 
The objective function is given by 
\begin{equation} \label{eq:J-pi}
\begin{split}
    J(t,x;\pi) & = \mathbb{E}_{t,x} \left[X_T^\pi + \lambda \int_t^T H(\pi_s) ds \right] - \frac{\gamma}{2} Var_{t,x} (X_T^\pi) \\
    & = \mathbb{E}_{t,x} \left[X_T^\pi - \frac{\gamma}{2} (X_T^\pi)^2 + \lambda \int_t^T H(\pi_s) ds \right] + \frac{\gamma}{2} \left(\mathbb{E}_{t,x} [X_T^\pi] \right)^2,
\end{split}
\end{equation}
where $\lambda > 0$ is the exploration weight and $\gamma > 0$ is the risk aversion factor.  
Additionally, $\mathbb{E}_{t,x}(\cdot), Var_{t,x}(\cdot)$ respectively denote the expectation and variance conditioned on $X_t^\pi = x$, i.e., $\mathbb{E}(\cdot | X_t^\pi = x)$ and $Var(\cdot | X_t^\pi = x)$. This means that the expectation is taken under the condition that the portfolio value process $X_t^\pi$ starts from $x$ at time $t$, following a given policy $\pi$.
Hence, the Exploratory Mean-Variance with Jumps (EMVJ) portfolio optimization problem is
\begin{equation} \label{problem-EMVJ}
    \sup_{\pi \in \mathcal{A}} J(t,x;\pi).
\end{equation}
We point out here that following an admissible policy $\pi \in \mathcal{A}$ as mentioned in Definition \ref{def:admissible-policy} also ensures $J(t,x;\pi) < \infty$. 

\begin{remark}
    We note that the entropy regularization term in the objective function (\ref{eq:J-pi}) plays a critical role in two major aspects. First, maximizing the classical mean-variance objective $\mathbb{E}_{t,x}[X_T^\pi] - \frac{\gamma}{2} Var_{t,x}(X_T^\pi)$ with an additional entropy term enforces the exploratory policy $\pi$ to avoid collapsing too quickly to a single deterministic control. This balances exploitation (picking known highly rewarding controls) with exploration (keeping the policy spread out), enabling possible discovery of outperforming controls that used to be unknown to the investor. More practically, the inclusion of an exploration weight $\lambda$ also allows us to control the extent to which an investor explores in the control space. 
    Second, the entropy term acts as a convex regularizer and yields optimal policies that are Boltzmann-like \citep{wang2020reinforcement}. It not only smooths the optimization landscape but also produces stochastic policy distributions (often Gaussian in Linear-Quadratic settings such as under our problem setup) that are easier to learn and evaluate. 
\end{remark}

The inclusion of the variance term \footnote{Recall that $Var_{t,x}(X) = \mathbb{E}_{t,x}(X^2) - (\mathbb{E}_{t,x}(X))^2$.} in the objective function (\ref{eq:J-pi}) introduces time-inconsistency, because it implies a nonlinear dependence on the conditional distribution of future wealth, violating Bellman’s principle of optimality \citep{bjork2021time}. 
When the investor re-optimizes at a later time with updated wealth and conditional expectation, her objective changes because the square of a conditional expectation introduces nonseparability over time. 
This leads to preference reversal, meaning that an investor who initially commits to an optimal strategy would deviate from it later in time once the optimization is recomputed. To address this, we can replace the classical notion of optimality with a {\it game-theoretic equilibrium control}, also known as a {\it time-consistent equilibrium control}, obtained by local (rather than global) optimality with respect to unilateral deviations in infinitesimal time. It is called an ``equilibrium” because the investment strategy is characterized as a subgame perfect Nash equilibrium of an intra-personal game between the investor’s present and future selves. In this sense, under the equilibrium control, each future self has no incentive to deviate, though it is generally not globally optimal.
We next introduce a formal definition of an {\it equilibrium investment policy.}

\begin{definition} \label{def:equilibrium-policy}
Let $\pi^{\ast }\in \mathcal{A}$ be an admissible investment
policy. For any $(t,x)\in \lbrack 0,T]\times \mathbb{R}$, consider a perturbed policy $\pi ^{\varepsilon }:=\{\pi_{s,y}^{\varepsilon }(\cdot )\}_{(s,y)\in \lbrack t,T]\times\mathbb{R}}$, defined as 
\begin{equation*}
\pi _{s,y}^{\varepsilon }(u),=%
\begin{cases}
\upsilon (u), & (s,y)\in [t,t+\varepsilon) \times \mathbb{R}, \\ 
\pi _{s,y}^{\ast }(u), & (s,y)\in [t+\varepsilon ,T]\times \mathbb{R},%
\end{cases}%
\end{equation*}%
where $\varepsilon >0$, and $\upsilon (\cdot )\in \mathcal{P}(\mathbb{R})$ is an arbitrary density function satisfying 
\begin{equation*}
\int_{\mathbb{R}} u^{2} \upsilon (u)du<\infty
\quad \text{and}\quad \int_{\mathbb{R}} \upsilon
(u)|\log \upsilon (u)| du < \infty,
\end{equation*}%
which implies $\pi ^{\varepsilon }\in \mathcal{A}$. Suppose that for any $%
(t,x)\in \lbrack 0,T]\times \mathbb{R}_{+}$ and any such perturbed strategy $%
\pi ^{\varepsilon }$, we have 
\begin{equation*}
\limsup_{\varepsilon \downarrow 0}\frac{J(t,x; \pi^\varepsilon)-J(t,x; \pi^\ast)}{\varepsilon }\leq 0.
\end{equation*}%
Then, $\pi ^{\ast }$ is an equilibrium investment policy for
problem~(\ref{problem-EMVJ}), and $J(t,x;\pi^{\ast})$ is the corresponding value
function.
\end{definition}

Define the {\it infinitesimal generator} for $\psi(t,x) \in C^{1,2}([0,T] \times \mathbb{R})$ with $\sup_{x}  \{\psi_{xx}(t,x) \} < \infty$ and $\pi \in \mathcal{A}$ as 

\begin{align}
    \mathcal{L}^\pi \psi(t,x) & = \int_\mathbb{R} \left[ u (\mu-r) \psi_x  + \frac{1}{2} \sigma^2 u^2 \psi_{xx} \right] \pi_{t,x}(u) du \notag \\
    & + \int_{\mathbb{R} \times [0,1]} [\psi(t, x + G^{\pi_{t,x}}(v) (e^z-1)) - \psi(t,x) - G^{\pi_{t,x}}(v) (e^z-1)\psi_x] \nu(dz) dv  \\
    & = \int_\mathbb{R} \left[ u (\mu-r) \psi_x  + \frac{1}{2} \sigma^2 u^2 \psi_{xx} 
    + \int_{\mathbb{R}} [\psi(t, x + u(e^z-1)) - \psi(t,x) - u(e^z-1)\psi_x] \nu(dz) \right] \pi_{t,x}(u) du, \label{eq:generator}
\end{align} 
where the last equality is given by the construction of $u = G^{\pi_{t,x}}(v)$ and thereby the fact that
\begin{align*}
    \int_{\mathbb{R}\times[0,1]} f(t,x,G^{\pi_{t,x}}(v), z) \nu(dz)dv = \int_{\mathbb{R}} \int_{\mathbb{R}} f(t,x,u,z) \nu(dz) \pi_{t,x}(u) du,
\end{align*}
for some function $f$. 


The next two theorems establish the theoretical core of this paper, which are critical foundations to design the RL algorithms in the subsequent section. 

\begin{theorem}[Verification Theorem] \label{thm:verification}
    Suppose that there exist two functions $V(t,x)$ and $g(t,x) \in \mathbb{C}^{1,2}([0,T]\times\mathbb{R})$ such that the following conditions are satisfied.
\begin{enumerate}
    \item[(i)] For any $(t,x) \in [0,T] \times \mathbb{R}$, 
    \begin{equation} \label{eq:HJB-V}
        \sup_{\pi \in\mathcal{A}} \left\{(\partial_t + \mathcal{L}^\pi)V + \lambda H(\pi) + \gamma g^\pi \mathcal{L}^\pi g^\pi - \frac{\gamma}{2} \mathcal{L}^\pi (g^\pi)^2 
        \right\} =0. 
    \end{equation}
    \item[(ii)] Let $\pi^*_{t,x}(u)$ be the maximizer that achieves the supremum in~(\ref{eq:HJB-V}), and suppose it satisfies, for any $(t,x) \in [0,T] \times \mathbb{R}$,
    \begin{equation} \label{eq:HJB-g}
        (\partial_t + \mathcal{L}^\pi) g^\pi = 0.
    \end{equation}
    \item[(iii)] For any $x \in \mathbb{R}$ and $\pi \in \mathcal{A}$, 
    \begin{equation} \label{eq:HJB-terminal}
        V(T,x) = g(T,x;\pi) = x.
    \end{equation}
\end{enumerate}

    Then $\pi^* := \{\pi_{t,x}(\cdot)\}_{(t,x)\in[0,T]\times\mathbb{R}}$ is an equilibrium policy for problem~(\ref{problem-EMVJ}). Moreover, we have the corresponding value function 
    \begin{equation} \label{eq:V=J-pi*}
        V(t,x) = J(t,x;\pi^*),
    \end{equation}
    and the auxiliary value function
    \begin{equation} \label{eq:g-pi*}
        g^{\pi^*}(t,x) \equiv g(t,x;\pi^*) := \mathbb{E}_{t,x} (X_T^{\pi^*}).    
    \end{equation}
\end{theorem}

\begin{theorem}[Equilibrium Policy and Value Functions] \label{thm:equilibrium-policy}
    An equilibrium investment policy for problem~(\ref{problem-EMVJ}) is given by
    \begin{equation} \label{eq:pi*}
        \pi^*_{t,x}(u) \sim N\left(\frac{\mu-r}{\gamma (\sigma^2 + \delta^2)} ,  \frac{\lambda}{\gamma (\sigma^2 + \delta^2)} \right), (t,x) \in [0,T]\times \mathbb{R}
    \end{equation}
    where $\delta^2 = \int_\mathbb{R} (e^z-1)^2 \nu(dz)$. 
    
    The corresponding value function is $V(t,x) = x + C(t)$ and the corresponding auxiliary value function is $g(t,x;\pi^*) = x + h^*(t)$, where
    \begin{align}
        C(t) & = (T-t)\left[\frac{(\mu-r)^2}{2\gamma(\sigma^2+\delta^2)} + \frac{\lambda}{2} \log \left(\frac{2\pi\lambda}{\gamma(\sigma^2+\delta^2)}\right) \right], \label{eq:C-pi} \\
        h^*(t) & = (T-t) \left[\frac{(\mu-r)^2}{\gamma(\sigma^2 + \delta^2)} \right] \label{eq:h-pi} .
    \end{align}
\end{theorem}

The proofs of Theorems \ref{thm:verification} and \ref{thm:equilibrium-policy} are given in Appendices \ref{apdx:proof-of-verification-theorem} and \ref{apdx:proof-of-equilibrium-policy-theorem} respectively.

\section{RL Algorithm} \label{sec:algorithm}
In this section, we design an RL algorithm to iteratively learn the equilibrium investment policy without the knowledge of the market and jump dynamics. We leverage the martingale property, in particular the {\it Orthogonality Condition}, from the analytical results in Theorems \ref{thm:verification} and \ref{thm:equilibrium-policy} to define a loss function, which we use to optimize our RL model. 

Recall that the stock price dynamics are:
\begin{equation}
    \frac{dS_t}{S_{t-}} = \mu_{true} dt + \sigma_{true} dW_t + \int_\mathbb{R} (e^z-1) \Tilde{N}(dt,dz),
\end{equation}
where $(\mu_{true} \in \mathbb{R}, \sigma_{true} > 0)$ are the true mean and volatility of stock returns.
We consider Merton's jump-diffusion model, which was studied by \cite{gao2024reinforcement} (see also \citep{merton1976option, bates1991crash, das2002surprise}). 
The jump density is 
\begin{equation}
    \nu(dz) = \zeta_{J, true} \frac{1}{\sqrt{2\pi\sigma_{J,true}^2}} \exp \left(-\frac{(z-\mu_{J,true})^2}{2\sigma_{J,true}^2}\right) dz, 
\end{equation}
where $\zeta_{J, true} > 0$ is the true arrival rate of jumps per unit time, and $(\mu_{J,true} \in \mathbb{R}, \sigma_{J,true} > 0)$ are the true mean and standard deviation of jump sizes. Most importantly, they are unknown parameters. 

Moreover, we note from~(\ref{eq:pi*}) that the equilibrium investment policy depends on $\mu, \sigma$,  and another parameter $\delta$, with 
\begin{equation} \label{eq:delta_true^2}
\begin{split}
    \delta_{true}^2 & := \int_\mathbb{R} (e^z-1)^2 \nu(dz)  \\
    & = \zeta_{J, true} \left[\exp(2\mu_{J,true} + 2\sigma^2_{J,true}) - 2\exp(\mu_{J,true} + \frac{1}{2}\sigma^2_{J,true}) + 1\right].
\end{split}
\end{equation}
We aggregate the jump parameters $(\zeta_J, \mu_J, \sigma_J)$ into $\delta$, and define the parameterization $\theta := (\mu, \sigma, \delta)$. The parameterized policy is $\pi^\theta := \{\pi^\theta_{t,x}(\cdot)\}_{(t,x)\in[0,T]\times \mathbb{R}}$, with
\begin{equation} \label{eq:pi-theta}
    \pi^\theta_{t,x}(u) \sim N\left(\frac{\mu-r}{\gamma (\sigma^2 + \delta^2)} ,  \frac{\lambda}{\gamma (\sigma^2 + \delta^2)} \right), (t,x) \in [0,T]\times \mathbb{R}.
\end{equation}
Under $\pi^\theta$, the portfolio value process in (\ref{eq:dX-pi}) becomes 
\begin{equation}  \label{eq:dX-theta}
    \begin{split} 
        dX_t^\theta = & \frac{(\mu_{true}-r)(\mu-r)}{\gamma(\sigma^2 + \delta^2)}dt + \sqrt{\sigma_{true}^2 \left(\frac{\lambda}{\gamma (\sigma^2 + \delta^2)} + \frac{(\mu-r)^2}{\gamma^2 (\sigma^2 + \delta^2)^2} \right)} dW_t \\
        & + \left(\frac{\mu-r}{\gamma (\sigma^2 + \delta^2)}\right) \int_\mathbb{R} (e^z - 1) \Tilde{N}(dt,dz).
    \end{split}
\end{equation}

Inspired by the HJB (\ref{eq:HJB-V}), we define, for a parameterized policy $\pi^\theta \in\mathcal{A}$, a stochastic process $M^\theta := \{M_t^\theta:= M(t, X_t^\theta; \pi^\theta)\}_{(t, X_t^\theta)\in[0,T]\times \mathbb{R}}$ such that $\forall (t,x) \in[0,T]\times \mathbb{R}$,
\begin{equation} \label{eq:M-theta}
\begin{split}
    M_t^\theta & = V^\theta(t, X_t^\theta) - \frac{\gamma}{2} (g^{\theta}(t, X_t^\theta))^2 + \lambda \int_0^t H(\pi^\theta_s) ds,
\end{split} 
\end{equation}
where $V^\theta(t,x) = x + C^\theta(t)$ and $g^\theta(t,x) = x + h^\theta(t)$ are the value function and the auxiliary value function corresponding to the parameterized policy $\pi^\theta$, respectively. 
Here, $C^{\theta}$ and $h^\theta$ are respectively given by~(\ref{eq:C-pi}) and (\ref{eq:h-pi}), with $\pi$ replaced by $\pi^\theta$.
\begin{theorem} \label{thm:M-is-martingale}
    The process $M^\theta$ is an $(\{\mathcal{F}_t\}_{t\in[0,T]}, \mathbb{P})$ martingale when $\theta = \theta_{true} := (\mu_{true}, \sigma_{true}, \delta_{true})$.
\end{theorem}
The proof of this theorem is postponed to Appendix \ref{apdx:proof-M-is-martingale}. 

Additionally, we know that any square-integrable martingale $M:=\{M_{t}\}_{t\in
\lbrack 0,T]}$ has an orthogonality condition 
\begin{equation*}
\mathbb{E}\left[ \int_{0}^{T}\xi _{t}dM_{t}\right] =0,
\end{equation*}%
where the test function $\xi :=\{\xi_{t}\}_{t\in \lbrack 0,T]}$ is an
arbitrary $\{\mathcal{F}_{t}\}_{t\in \lbrack 0,T]}$-adapted process that is
square-integrable with respect to $M$. This, together with the martingale
property of $M^{\theta }$, allows us to define an {\it Orthogonality Condition (OC) Loss} function. Following \cite{PE}, we choose the test function
as the partial derivative of the parameterized value function $V^{\theta }$
with respect to its parameters $\theta $.

\begin{definition}[Orthogonality Condition (OC) Loss] 
Take $\theta = (\mu, \sigma, \delta) = (\theta_1, \theta_2, \theta_3)$. 
For $(t,x)\in [0,T]\times \mathbb{R}$ and $j=1,2,3$, the OC
loss is given by 
\begin{equation}
    L_{OC}(\theta_j) = \mathbb{E}_{t,x} \left[ \int_0^T \frac{\partial V^\theta}{\partial \theta_j}(t, X_t^{\pi^\theta}) d M_t^\theta  \right],  
\end{equation}
with 
\begin{align}
    \frac{\partial V^\theta}{\partial \theta_1}(t,x) & = \frac{\partial V^\theta}{\partial \mu} = (T-t)\left[\frac{(\mu-r)}{\gamma(\sigma^2+\delta^2)} + \frac{\lambda}{2} \log \left(\frac{2\pi\lambda}{\gamma(\sigma^2+\delta^2)}\right) \right] \label{eq:dV_dmu}, \\
    \frac{\partial V^\theta}{\partial \theta_2}(t,x) & = \frac{\partial V^\theta}{\partial \sigma} = -(T-t) \left[\frac{(\mu-r)^2\sigma}{\gamma(\sigma^2 + \delta^2)^2} + \frac{\lambda \sigma}{(\sigma^2 + \delta^2)}\right] \label{eq:dV_dsigma}, \\
    \frac{\partial V^\theta}{\partial \theta_3}(t,x) & = \frac{\partial V^\theta}{\partial \delta} = -(T-t) \left[\frac{(\mu-r)^2\delta}{\gamma(\sigma^2 + \delta^2)^2} + \frac{\lambda \delta}{(\sigma^2 + \delta^2)}\right]. \label{eq:dV_dtilde_sigma} 
\end{align}
\end{definition}

\begin{remark}
    The orthogonality condition loss, first introduced by \cite{PE}, is built on the fact that under the true value function, the temporal difference residual (obtained from It\^o’s formula and the HJB equation) must be a martingale with zero drift, hence orthogonal to every square-integrable test function. 
    Intuitively, if $V^\theta$ correctly approximates the true value function, i.e., $\theta = \theta_{true}$, then no predictable component should remain in the temporal difference residual, 
    resulting in zero conditional expectation.    
    Note that under our RL setting, the parameterized policy $\pi^\theta$ and the corresponding value function $V^\theta$ share the same parameters, thus the policy optimization and the value function approximation are achieved simultaneously when $\theta = \theta_{true}$.  
    Leveraging the equivalence of parameter convergence and the martingale property from Theorem \ref{thm:M-is-martingale}, 
    the orthogonality condition loss enforces effective updates of policy and value functions, by iteratively updating $\theta$ and reducing the loss to zero. 
\end{remark}

Considering a discretization $0 = t_0 < \cdots < t_N = T$ with mesh size $\Delta t$, the discretized loss function becomes
\begin{equation} \label{eq:OC-loss-discrete}
\begin{split}
    L_{OC}(\theta_j) = & \sum_{n=0}^{N-1} \frac{\partial V^\theta}{\partial \theta_j} (t_n, X_{t_n}^{\pi^\theta}) 
    \Biggl\{ V^\theta(t_{n+1}, X_{t_{n+1}}^{\pi^\theta}) - V^\theta(t_n, X_{t_n}^{\pi^\theta}) \\
    & - \frac{\gamma}{2} 
    \left[g^\theta(t_{n+1}, X_{t_{n+1}}^{\pi^\theta})^2 - g^\theta(t_{n}, X_{t_{n}}^{\pi^\theta})^2 \right] + \frac{\lambda}{2} \log \left(\frac{2\pi e \lambda}{\gamma(\sigma^2 + \delta^2)}\right) \Delta t
    \Biggr\}.  
\end{split}
\end{equation}

\bigskip

We summarize the whole training procedure in Algorithm \ref{alg}. 

\begin{algorithm}[H]
\SetAlgoLined
Initialize the number of epochs $N_{epochs}$, the investment horizon $T$, the mesh size of the continuous time discretization $\Delta t$, the exploration parameter $\lambda > 0$, the initial portfolio value $x_0 > 0$, the learning rates $\eta = (\eta_{1}, \eta_{2}, \eta_{3})$, the ``grounding true" market parameters $\theta_{true} = (\theta_{true,1}, \theta_{true,2}, \theta_{true,3})$ $\equiv (\mu_{true}, \sigma_{true}, \delta_{true})$, and the model parameters $\theta^{(0)} = (\theta_1^{(0)}, \theta_2^{(0)}, \theta_3^{(0)}) = (\mu^{(0)}, \sigma^{(0)}, \delta^{(0)})$.

Compute $N = \frac{T}{\Delta t}$. 

\For{$k = 0, \cdots, N_{epoch}-1$}{
Fix a realized path of Brownian motion $\{W_{t_n}^{(k)}\}_{n=0}^N$, with $\Delta W_{t_n}^{(k)} := W^{(k)}_{t_{n+1}} - W^{(k)}_{t_n} \sim N(0, \Delta t)$. 

Sample number of jumps $N_k \sim Poisson(\zeta_{J,true} \Delta t)$ and sample the jumps $(Z_1^{(k)}, \cdots, Z_{N_k}^{(k)}) \overset{iid}{\sim} N(\mu_{J,true}, \sigma_{J,true}^2)$. 

Simulate $\{X^{(k)}_{t_n}\}_{n=0}^N$ via (\ref{eq:dX-theta}), with $(\mu,\sigma, \delta)$ replaced by $(\mu^{(k)}, \sigma^{(k)}, \delta^{(k)})$, $\{W_{t_n}\}_{n=0}^N$ replaced by $\{W_{t_n}^{(k)}\}_{n=0}^N$, and the integral replaced by $\sum_{j=1}^{N_k} (e^{Z_j^{(k)}} - 1)$.

Compute $\{\frac{\partial V^{\theta^{(k)}}}{\partial \mu}(t_n, X_{t_n}^{(k)})\}_{n=0}^N$, $\{\frac{\partial V^{\theta^{(k)}}}{\partial \sigma}(t_n, X_{t_n}^{(k)})\}_{n=0}^N$ and $\{\frac{\partial V^{\theta^{(k)}}}{\partial \delta}(t_n, X_{t_n}^{(k)})\}_{n=0}^N$ via (\ref{eq:dV_dmu}), (\ref{eq:dV_dsigma}) and (\ref{eq:dV_dtilde_sigma}), with $(\mu,\sigma, \delta)$ replaced by $(\mu^{(k)}, \sigma^{(k)}, \delta^{(k)})$.

\For{$j = 1, 2, 3$}{
Compute $L(\theta_j^{(k)}) := L_{oc}(\theta_j^{(k)}; \{X_{t_n}^{(k)}\}_{n=0}^N)$ via (\ref{eq:OC-loss-discrete}).


Update $\theta_j^{(k+1)} \leftarrow \theta_j^{(k)} + \eta_j \times L(\theta_j^{(k)})$
}
}
\caption{Training with Orthogonality Condition Loss} \label{alg}
\end{algorithm}

\section{Numerical Results} \label{sec:numerical}
In this section, we train and evaluate our model on simulated data and real market data. On simulated data (Subsection \ref{sec:numerical-simulated}), we train our model and show that the model parameters converge to the corresponding true values. Moreover, the investment performance (in terms of mean terminal portfolio value) on the simulated data is close to the theoretical expectation. We also examine our model on real market data (Subsection \ref{sec:numerical-real}), which illustrates the potential profitability of our equilibrium policy.

\subsection{Analysis On Simulated Market Data} \label{sec:numerical-simulated}
From \href{https://finance.yahoo.com/quote/\%5ESP500TR/}{Yahoo Finance}, we collect the S\&P500 market index (total return) as a ``representative" stock at a daily frequency, ranging from January 1st of 2000 to December 31st of 2023. 
We fit Merton's jump-diffusion model, parameterized by both the market parameters $(\mu, \sigma)$ and the jump parameters $(\mu_J, \sigma_J, \zeta_J)$. 
Denote by $\theta := (\mu, \sigma, \mu_J, \sigma_J, \zeta_J)$. Consider a time discretization $0 = t_0 < \cdots < t_N = T$ with mesh size $\Delta t$. When $\Delta t$ is small, we suppose the log-returns $R_i := \log \left(\frac{S_{t_i}}{S_{t_{i-1}}}\right)$, for $i=1,\cdots,\frac{T}{\Delta t}$, are approximately independently and identically distributed with density function 
\begin{equation*}
    f(r;\theta) = \sum_{m=0}^\infty p_m(\zeta_J, \Delta t) \cdot \phi(r; \nu_m(\theta,\Delta t), \tau^2_m(\theta, \Delta t)), 
\end{equation*}
where $p_m(\zeta_J, \Delta t) = \frac{e^{-\zeta_J \Delta t} (\zeta_J \Delta t)^m}{m!}$ is the Poisson probability of $m$ jumps, $\phi(\cdot; \nu,\tau^2)$ is the Gaussian density with mean $\nu$ and variance $\tau^2$, and 
\begin{align*}
    \nu_m(\theta, \Delta t) & = \left(\mu-\frac{1}{2}\sigma^2 - \lambda (e^{\mu_J+\frac{1}{2}\sigma_J^2} - 1) \right) \Delta t + m \mu_J \\
    \tau^2_m(\theta, \Delta t) & = \sigma^2 \Delta t + \mu \sigma_J^2.
\end{align*}
While the direct computation of an infinite sum $f(r,\theta)$ is impractical, we approximate it with a truncated sum up to m=2. Because multiple jumps per interval are very rare and $p_m(\zeta_J, \Delta t)$ for $m > 2$ are negligible.  
We calculate the likelihood function as $L(\theta;R) = \prod_{i=1}^{T/\Delta t} f(R_i;\theta)$ and apply the Nelder-Mead method \citep{nelder1965simplex} to obtain the maximum likelihood estimates (MLE) of $\theta$. We choose the MLEs of $\theta$ as the ``grounding true" parameter values $(\mu_{true}, \sigma_{true}, \zeta_{J,true}, \mu_{J, true}, \sigma_{J, true})$ for simulation.
In this way, our simulated data reflects the real market. 

We consider a one-year investment horizon $(T=1)$ with daily rebalancing $(\Delta t = \frac{1}{252})$, reflecting an assumption of 252 trading days per year. The investor starts with an initial wealth of $x_0 = 1$, and invests in the stock following the equilibrium policy (\ref{eq:pi*}) with initial parametrization $(\mu_0, \sigma_0 ,\delta_0) = (0.1, 0.1, 0.05)$. 
For simplicity, we suppose there is no cost of borrowing, i.e., the risk free rate $r = 0$. The exploration weight $\lambda$ and the risk aversion parameter $\gamma$ are set as $1$ for training.
We summarize all the parameters in Table \ref{tab:params-on-sim}.  

\begin{table}[H]
    \centering
    \begin{tabular}{c|c}
    \toprule
        Parameters & Values \\
    \midrule
        $x_0$ & 1\\
        ($T, \Delta t)$ & $(1, \frac{1}{252})$\\
        $\lambda$ & 1 \\
        $\gamma$ & 1 \\
        $r$ &  0.01 \\
        $(\mu_{true}, \sigma_{true})$ & (0.0878, 0.1321)\\
        $\zeta_{J,true}$ &  27.6813 \\
        $(\mu_{J,true}, \sigma_{J,true})$ &  (-0.0040, 0.0274)\\
        $\delta_{true}$ (by (\ref{eq:delta_true^2})) &  0.1449 \\
        $\delta_{0}$ & 0.05 \\
        $(\mu_0, \sigma_0)$ & (0.1, 0.1)  \\
    \bottomrule
    \end{tabular}
    \caption{Parameters for training and evaluation on simulated data.}
    \label{tab:params-on-sim}
\end{table}

\begin{figure}[H]
    \centering
    \includegraphics[width=1\linewidth]{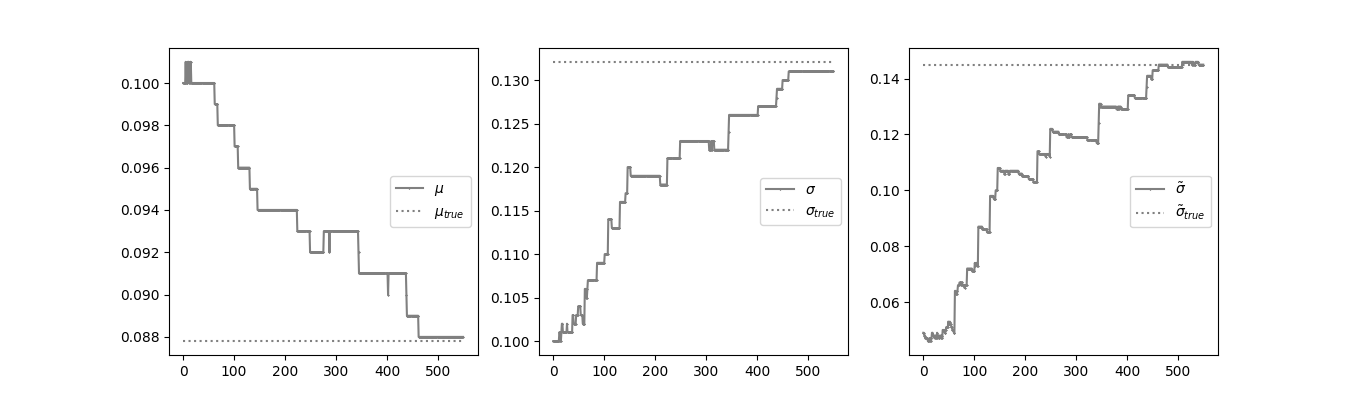}
    \caption{Convergence of $\mu$ (left), $\sigma$ (middle) and $\delta$ (right). }
    \label{fig:param-convergence}
\end{figure}

We train the model following Algorithm \ref{alg} and all the model parameters converge to their ``grounding true" values in 550 epochs as shown in Figure \ref{fig:param-convergence}. 
This indicates that our model has been optimized to the equilibrium policy. 
We make a note here that the three parameters do not converge at the same rate --- in the case showed in Figure \ref{fig:param-convergence}, $\mu$ converges fastest while $\sigma$ converges slowest \footnote{The speed of parameter convergence depends on how close the initial points are to the corresponding true values.}. 
In the training process, we impose a linear learning rate scheduler to control the learning rates for each parameter. 

Next, we check if our trained model is able to achieve the theoretically expected investment performance. From Theorem \ref{thm:equilibrium-policy}, we know that if an investor follows a policy $\pi^*$, the expected terminal portfolio value is 
\begin{equation*}
\begin{split}
    \mathbb{E}_{0, x_0}(X_T^{\pi^*}) & = g(t,x;\pi^*) = x_0 + h^{\pi^*}(t) \\
    & = x_0 + (T-t) \left[\frac{(\mu-r)^2}{\gamma(\sigma^2 + \delta^2)} \right].
\end{split}
\end{equation*}
We suppose the investor follows the mean of the equilibrium policy (\ref{eq:pi*}), that is, at each time $t$, she invests $u_t = \frac{\mu}{\gamma(\sigma^2+\delta^2)}$ dollars in the stock, and saves the remaining $X_t - u_t$ in cash at the risk-free interest rate $r$. 
If $u_t > X_t$, this means the investor is investing with leverage, and the cost of borrowing is equal to the risk-free interest rate $r$.
If $u_t < 0$, this means the investor is short-selling the stock, and saving the premium at the risk-free interest rate $r$. 
We highlight that investing at our policy mean is an extension to the equilibrium policy considered in \citep{bjork2021time}, as they coincide with each other when there are no jumps in the market dynamics. 
While we train the model with risk aversion parameter $\gamma = 1$, we examine the investment performance for investors with different risk appetites, $\gamma \in [0.1, 0.5, 1, 2, 5]$. Higher $\gamma$ values represent more risk averse investors, hence holding a smaller long or short position in the stock. 
For each investor, we simulate 100 stock price trajectories and make investments repeatedly. As a result, we get 100 portfolio value trajectories. The {\it realized mean} and {\it realized volatility} of the terminal portfolio value are calculated as the average and the standard deviation of the last entry over the 100 portfolio value trajectories. We also compute and contrast the {\it empirical objective function}, $J_{empirical}$ using (\ref{eq:J-pi}) with $\lambda = 0$ \footnote{Because we are following the mean of the equilibrium policy, there is no policy exploration.}, 
and the theoretical value of the objective function $V$ given the equilibrium policy, using the value function in Theorem \ref{thm:equilibrium-policy}). 

We summarize the investment performance in Table \ref{tab:params-on-sim}. For the realized mean, realized volatility and $J_{empirical}$, we also compute the Monte Carlo standard errors following the metrics in \citep{morris2019using}, which are reported in the parenthesis in Table \ref{tab:params-on-sim}. 
The realized mean terminal portfolio values are pretty close to the corresponding theoretical means. The realized volatility diminishes as the investor becomes more risk averse (i.e., as $\gamma$ increases), which is practically meaningful as more aggressive investment results in higher volatility. The empirical objective function values are reasonably close to the theoretical values, and the gap narrows as the investor becomes more risk averse. 
\begin{table}[H]
    \centering
    \begin{tabular}{c|c|c|c|c|c}
    \toprule 
        $\gamma$ & realized mean & theoretical mean & realized volatility & $J_{empirical}$ & theoretical $V$ \\
    \midrule 
        0.1 & 2.9980 (0.3032) & 3.0213 & 3.0316 (0.2154) & 2.5384 (0.3055) & 2.0106 \\
        0.5 & 1.3996 (0.0606) & 1.4043 & 0.6063 (0.0431) & 1.3077 (0.0611) & 1.2021 \\
        1 & 1.1998 (0.0303) & 1.2021 & 0.3032 (0.0215) & 1.1538 (0.0305) & 1.1011 \\
        2 & 1.0999 (0.0152) & 1.1011 & 0.1516 (0.0108) & 1.0770 (0.0153) & 1.0505 \\
        5 & 1.0400 (0.0061) & 1.0404 & 0.0606 (0.0043) & 1.0308 (0.0061) & 1.0202 \\
    \bottomrule
    \end{tabular}
    \caption{Investment performance on simulated market data. The Monte Carlo standard errors of the realized means, realized volatilities and $J_{empirical}$ are reported in the parenthesis.}
    \label{tab:performance-on-sim}
\end{table}



\subsection{Analysis on Real Market Data} \label{sec:numerical-real}

To evaluate our model on real market data, we choose the S\&P500 market index mentioned in the last subsection as the risky asset. We also collect the US 3-month Treasure Bill (3mTBill) from \href{https://finance.yahoo.com/quote/\%5EIRX/}{Yahoo Finance} as the risk free interest rate. The frequency of the 3mTBill data is daily and the range is from January 1st, 2000 to December 31st, 2023, which matches the S\&P500 data. 

For model training, we divide the investment horizon of 24 years, from 2000 to 2023, into 14 eleven-year rolling windows, 2000-2010, 2001-2011, $\cdots$, 2013-2023. 
Each rolling window spans from the first day of the starting year to the last day of the ending year, e.g., window 2000-2010 covers the period from January 1st, 2000 to December 31st, 2010. The first 10 years of each window is the {\it training period} and the final year is the {\it evaluation period}. 
We suppose an investor is risk averse with $\gamma = 5$. 
In the training phase, we set $\lambda = 5$ to encourage exploration within the control space. Over each rolling window, we fit a jump-diffusion model on the training period and use the MLEs as the initial parameters of the RL model. We evaluate the model performance on the evaluation period of each rolling window. 

The investment performances on the training periods and the evaluation periods are given in Tables \ref{tab:performance-on-train} and \ref{tab:performance-on-eval}, respectively, including the terminal portfolio value means, volatilities and Sharpe Ratios (SR). 
Over each training or evaluation period, we simultaneously create 100 self-financing portfolios each with initial value of $x_0 = 1$. We rebalance the portfolios on a daily basis according to the stock and risk-free interest rate dynamics, following the equilibrium policy (\ref{eq:pi*}). To reduce the portfolio volatility, we set $\lambda = 0.01$ if we invest over the training periods and set $\lambda = 0.1$ if we invest over the evaluation periods. The {\it mean} and the {\it volatility} columns are calculated as the average and standard deviation of the 100 portfolios' values on the last day of the investment period. The Monte Carlo standard errors are reported in the parenthesis, following the metrics provided in \citep{morris2019using}. The {\it risk-free} column values are the terminal values if we hold the risk-free asset throughout the investment period. 

\begin{table}[H]
    \centering
    \begin{tabular}{c|c|c|c}
    \toprule
        window & mean & volatility & risk-free \\
    \midrule
        2000-2010 & 1.3980 (0.0198) & 0.1978 (0.0141) & 1.3071  \\
        2001-2011 & 1.3859 (0.0202) & 0.2022 (0.0144) & 1.2348  \\
        2002-2012 & 1.3845 (0.0185) & 0.1848 (0.0131) & 1.1944  \\
        2003-2013 & 1.8165 (0.0201) & 0.2007 (0.0143) & 1.1766  \\
        2004-2014 & 1.8445 (0.0187) & 0.1869 (0.0133) & 1.1654  \\
        2005-2015 & 1.9569 (0.0140) & 0.1397 (0.0099) & 1.1499  \\
        2006-2016 & 1.5993 (0.0150) & 0.1500 (0.0107) & 1.1148  \\
        2007-2017 & 1.5213 (0.0154) & 0.1535 (0.0109) & 1.0667  \\
        2008-2018 & 1.6454 (0.0146) & 0.1463 (0.0104) & 1.0311  \\
        2009-2019 & 3.6082 (0.0207) & 0.2066 (0.0147) & 1.0370  \\
        2010-2020 & 4.9670 (0.0195) & 0.1945 (0.0138) & 1.0570  \\
        2011-2021 & 5.5411 (0.0272) & 0.2720 (0.0193) & 1.0592  \\
        2012-2022 & 7.4664 (0.0231) & 0.2311 (0.0164) & 1.0591  \\
        2013-2023 & 5.2763 (0.0225) & 0.2246 (0.0160) & 1.0795  \\
    \bottomrule
    \end{tabular}
    \caption{Investment performance on training periods. The model was trained with $(\lambda, \gamma) = (5, 5)$ and the table is calculated with $(\lambda, \gamma) = (0.01, 5)$. The Monte Carlo standard errors of the portfolio value means and volatilities are reported in the parenthesis.}
    \label{tab:performance-on-train}
\end{table}

Clearly, Table \ref{tab:performance-on-train} shows that all the portfolios invested over the training periods achieve higher mean returns than the risk-free return, with reasonably low volatility. 
The mean terminal portfolio values are very high during the post subprime crisis in 2008 --- the portfolio value triples in 2009-2019 and even grows sevenfold in 2012-2022, in spite of the breakout of COVID since 2020. 
Even in the training periods that are close to or cover the subprime crisis, the training performance is still reliable, as the mean terminal portfolio value grows at least by around 50\% in 2004-2014, 2005-2015, 2006-2016, 2007-2017 and 2008-2018. 
This implies that our models are well-trained and the market parameters of the training periods are properly learned through the training process.

\begin{figure}[H]
    \centering
    \includegraphics[width=1\linewidth]{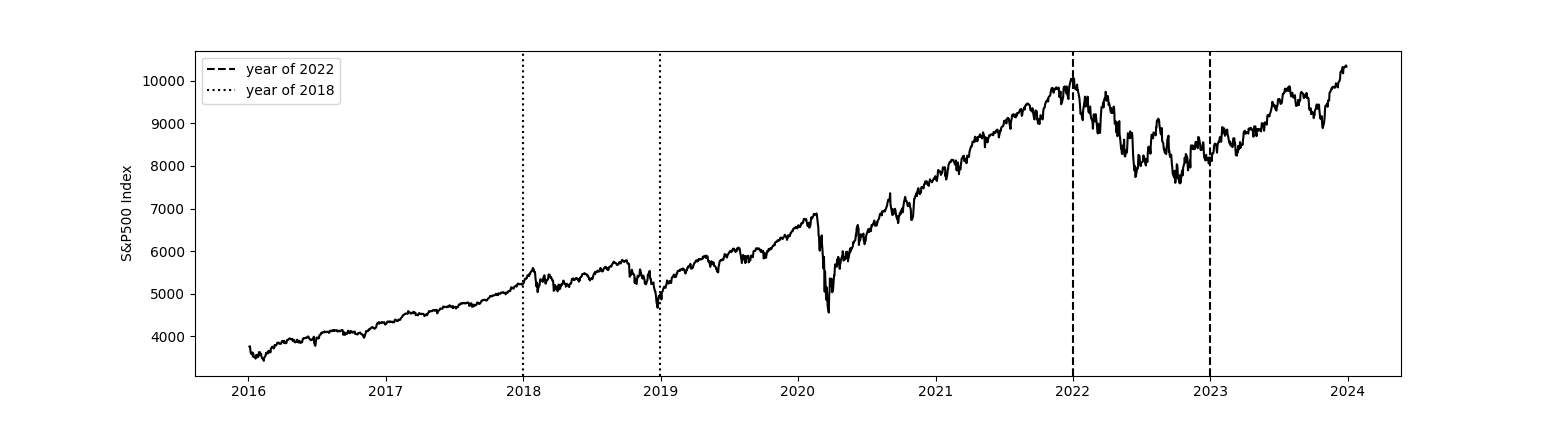}
    \caption{S\&P500 market index. The period between the dotted vertical lines is the year 2018 and the period between the dashed vertical lines is the year 2022. The market was significantly shocked due to the crash of \href{https://en.wikipedia.org/wiki/Cryptocurrency_bubble\#2017_boom_and_2018_crash}{cryptocurrency bubble} in 2018, as well as the post-COVID historical \href{https://www.reuters.com/markets/global-central-banks-deliver-historic-rate-hike-blast-2022-2022-12-23/}{global interest rate hike} and the escalation in \href{https://en.wikipedia.org/wiki/Russo-Ukrainian_war_(2022–present)}{the Russo-Ukrainian war} in 2022.}
    \label{fig:SP500}
\end{figure}

Table \ref{tab:performance-on-eval} conveys the generalizability and practical applicability of our model. In 12 out of the 14 evaluation periods, the mean terminal portfolio values are higher than the risk-free portfolio, with reasonably low volatility. 
The only two exceptions are in 2018 and 2022, during which the market trend flips, as shown in Figure \ref{fig:SP500}. That is, the market trend of the training period (2008-2017 and 2012-2021) is opposite to that of the evaluation period (2018 and 2022, respectively). So, the model is mislead by the training period. For example, in the window 2012-2022, the model is trained over a bullish period of 2012-2021 and it will assume the market to continue growing in 2022, which is not the case due to the breakout of COVID. As a result, the model decides to hold a long position in the stock during a bearish evaluation period in 2022. 
However, in the windows 2009-2019 and 2013-2023, our model takes the years of 2018 and 2022 as part of the training periods and adjusts the investment strategies, leading to an improved performance in the evaluation periods in 2019 and 2023. 
This implies that our model is robust against different market conditions, as long as the market does not significantly violate the model assumptions. Moreover, our model preserves the potential of reversing the adverse performance, by gradually digesting the market dynamics and adjusting the investment strategies accordingly. 

For purposes of comparison, we also present the investment performance using the MLEs of a fitted Merton jump-diffusion model as the parameters of the equilibrium investment policy. We use the subscript $MLE$ to denote the investment performances using MLEs in Table \ref{tab:performance-on-eval}. Except for 2018 and 2022, our model achieves higher Sharpe Ratios in all other evaluation periods, compared to the investment strategy using MLEs. Since our model takes the MLEs as the initial parameters in the training phase, the improved investment performance demonstrates the effectiveness of our algorithm. In the years 2018 and 2022, the investments using MLEs also obtain negative Sharpe Ratios, which provides a poor starting point for us to train the model. This partially explains why our model performance in 2018 and 2022 is poor.

\begin{table}[H]
    \centering
    \begin{small}
    \begin{tabular}{c|c c c|c|c c c}
    \toprule
        year & mean & volatility & SR & risk-free & $mean_{MLE}$ & $volatility_{MLE}$ & $SR_{MLE}$ \\
    \midrule
        2010 & 1.0666 (0.0173) & 0.1726 (0.0123) & 0.3784 & 1.0013 & 1.0189 (0.0121) & 0.1214 (0.0086) & 0.1452  \\
        2011 & 1.0227 (0.0186) & 0.1863 (0.0132) & 0.1197 & 1.0004 & 1.0067 (0.0129)  & 0.1292 (0.0092) & 0.0487  \\
        2012 & 1.0621 (0.0105) & 0.1054 (0.0075) & 0.5820 & 1.0008 & 1.0226 (0.0081) & 0.0813 (0.0058) & 0.2677  \\
        2013 & 1.2297 (0.0101) & 0.1006 (0.0071) & 2.2791 & 1.0005 & 1.1121 (0.0077) & 0.0774 (0.0055) & 1.4414  \\
        2014 & 1.2132 (0.0124) & 0.1244 (0.0088) & 1.7127 & 1.0002 & 1.0481 (0.0077) & 0.0773 (0.0055) & 0.6187  \\
        2015 & 1.0372 (0.0170) & 0.1696 (0.0121) & 0.2171 & 1.0004 & 1.0095 (0.0108) & 0.1083 (0.0077) & 0.0837  \\
        2016 & 1.0625 (0.0099) & 0.0985 (0.0070) & 0.6029 & 1.0030 & 1.0457 (0.0089) & 0.0894 (0.0064) & 0.4773  \\
        2017 & 1.0880 (0.0043) & 0.0427 (0.0030) & 1.8451 & 1.0093 & 1.0752 (0.0040) & 0.0404 (0.0029) & 1.6309  \\
        2018 & 0.9975 (0.0141) & 0.1412 (0.0100) & -0.1548 & 1.0194 & 1.0098 (0.0118) & 0.1184 (0.0084) & -0.0814  \\
        2019 & 1.4672 (0.0138) & 0.1384 (0.0098) & 3.2274 & 1.0206 & 1.2435 (0.0107) & 0.1069 (0.0076) & 2.0840  \\
        2020 & 1.6298 (0.0476) & 0.4761 (0.0338) & 1.3158 & 1.0034 & 1.2601 (0.0345) & 0.3449 (0.0245) & 0.7445  \\
        2021 & 1.8431 (0.0173) & 0.1726 (0.0123) & 4.8834 & 1.0003 & 1.2414 (0.0108) & 0.1075 (0.0076) & 2.2434  \\
        2022 & 0.2339 (0.0356) & 0.3555 (0.0252) & -2.2119 & 1.0202 & 0.7706 (0.0225) & 0.2246 (0.0160) & -1.1109 \\
        2023 & 1.5263 (0.0201) & 0.2063 (0.0147) & 2.3002 & 1.0517 & 1.1563 (0.0124) & 0.1239 (0.0088) & 0.8440 \\
    \bottomrule
    \end{tabular}
    \end{small}
    \caption{Investment performance on evaluation periods. The model was trained with $(\lambda, \gamma) = (5, 5)$ and the table is calculated with $(\lambda, \gamma) = (0.1, 5)$. SR denotes the Sharpe Ratio, and the subscript $MLE$ denotes the investment performance calculated using MLEs of a Merton's jump-diffusion model. The Monte Carlo standard errors of the portfolio value means and volatilities are reported in the parenthesis.}
    \label{tab:performance-on-eval}
\end{table}

Finally, we also consider a more conservative investor with $\lambda = 1$. After retraining and retesting the model, we improve the investment performance in 2018, as shown in Table \ref{tab:performance-on-eval-2018}. This reiterates our model's robustness against violation of model assumptions in the real-world market. 

\begin{table}[H]
    \centering
    \begin{tabular}{c|c|c|c}
    \toprule
        year & mean & volatility & SR \\
    \midrule
        2018 & 1.0598 (0.0151) & 0.1507 (0.0107) & 0.2679 \\
    \bottomrule
    \end{tabular}
    \caption{Investment performance in 2018. The model was retrained with $(\lambda, \gamma) = (1, 5)$ and the table is calculated with $(\lambda, \gamma) = (0.01, 5)$. SR denotes the Sharpe Ratio. The Monte Carlo standard errors of the portfolio value mean and volatility are reported in the parenthesis.}
    \label{tab:performance-on-eval-2018}
\end{table}

\section{Conclusion} \label{sec:conclusion}

In this paper, we apply the time-inconsistent control (TIC) approach to a RL-facilitated exploratory Mean-Variance problem with jump-diffusion market dynamics. Unlike many past works which choose to solve the Lagrangian dual of the MV problem \citep{wang2020continuous, gao2024reinforcement, wu2024reinforcement, chen2025EMVRS, gao2024reinforcement}, we tackle the time-inconsistency of the MV problem following the methods introduced in \citet{bjork2021time}. We acknowledge that the uniqueness of an equilibrium solution to a TIC problem remains as an open research topic, as discussed in \citet{dai2023learning}. Therefore, we disregard the uniqueness argument of our solution, and decide to present one solution by establishing and solving the extended HJB equations. We present a verification theorem (Theorem \ref{thm:verification}) that ensures our solution is an equilibrium solution, according to Definition \ref{def:equilibrium-policy}. 

Furthermore, we choose the RL model parameterization that is practically meaningful, based on our analytical solutions in Theorem \ref{thm:equilibrium-policy}. We then leverage the martingale property, specifically the orthogonality condition, to define the Orthogonality Condition (OC) loss function, which is more effective in our problem context compared to another widely-applied Temporal-Difference (TD) loss in RL, as discussed in \citep{PE, chen2025EMVRS}. With our choice of parameterization and loss function, we are able to check whether parameters can converge to the corresponding true values. 
Thankfully, all parameters converge correctly in our simulation study, indicating that our RL algorithm is effective to learn the equilibrium policy for jump-diffusion market dynamics. 

On real market data, we examine our model over a course of 24 years since 2000. Our model performs well on all the training periods and is profitable in 13 out of 14 evaluation periods, with reasonably low volatility. We acknowledge that our model can be mislead if the market significantly violates the model assumptions.

\bigskip
We conclude this paper with two potential directions for future work. 
\begin{itemize}
    \item[(i)] {\it Bayesian approach to learn the model parameters.} We acknowledge the possibility of learning our model parameters via a Bayesian approach, since our model parameters all have practical meanings and the number of parameters is small. One potential way is to apply a filtering method \citep{fulop2011filtering} to iteratively learn the posterior distributions of each model parameter. The stochasticity of the exploratory control thereby has two sources, one from the distribution-valued policy and the other from the posterior distribution of the model parameters. 

    Besides, one can also employ Bayesian RL \citep{vlassis2012bayesian, ghavamzadeh2015bayesian} techniques to learn the posterior distribution of the model parameters, instead of learning a single estimate as in the classical RL approaches. 
    
    \item[(ii)] {\it Extension to a regime-switching market.} Past work has considered portfolio optimization in a regime-switching and jump-diffusion market \citep{savku2022stochastic, zhang2021optimal, zhang2022portfolio, gal2005dynamic}. However, to the best of our knowledge, an exploratory formulation of this problem remains untouched.
    
    As we discussed in Section \ref{sec:numerical-real}, our model can perform poorly if the market trend in a training period is opposite to that in an evaluating period. Such a trend-flipping phenomenon can potentially be explained by a shift of market regimes. \citet{chen2025EMVRS} solved an exploratory MV problem in a regime-switching market, although their problem setting excludes jumps and they took a time-consistent approach. 
\end{itemize}

\newpage 

\printbibliography[title=Bibliography]

\newpage
\appendix
\section{Proofs}

\subsection{Proof of Lemma \ref{lm: admissible-policy-to-SDE-solution}} \label{apdx:proof-of-lemma-admissible-policy-to-SDE-solution}
Suppose $\pi \in \mathcal{A}$. By the fact that $Var(U) = E[U^2] - (E[U])^2 \geq 0$ for $U \sim \pi$, 
\begin{align*}
    \int_\mathbb{R} u^2 \pi_{t,x}(u) du \geq \left(\int_\mathbb{R} u \pi_{t,x}(u) du  \right)^2. 
\end{align*}
Furthermore, because $u = G^{\pi_{t,x}}(v)$ for $(t,x) \in [0,T] \times \mathbb{R}$ and hence $dv = \pi_{t,x}(u)du$, we have
\begin{align*}
    \int_0^1 |G^{\pi_{t,x}}(v)|^2 dv = \int_\mathbb{R} u^2 \pi_{t,x}(u)dv. 
\end{align*}

Then, by (2) of Definition \ref{def:admissible-policy} and the conditions on the L\'evy measure (\ref{eq:conditions-of-Levy-measure}), we know there exists a constant $C_4$ such that
\begin{align*}
    & \left(\int_\mathbb{R} u \pi_{t,x}(u) du  \right)^2 \leq C_4(1+x^2) \\
    & \int_{\mathbb{R} \times[0,1]} |G^{\pi_{t,x}}(v) (e^z-1)|^2 \nu(dz) dv = \left(\int_{\mathbb{R}} (e^z-1)^2 \nu(dz)\right) \times 
    \left(\int_0^1 |G^{\pi_{t,x}}(v)|^2 dv \right) 
    \leq C_4(1+x^2)
\end{align*}
for any $(t,x) \in [0,T] \times \mathbb{R}$.
Therefore, there exists $C' := \max(C1, 2 C4)$ such that 
\begin{equation*}
    \left(\int_\mathbb{R} u \pi_{t,x}(u) du  \right)^2
    + \int_\mathbb{R} u^2 \pi_{t,x}(u) du 
    + \int_{\mathbb{R} \times[0,1]} |G^{\pi_{t,x}}(v) (e^z-1)|^2 \nu(dz) du 
    \leq C'(1+x^2), 
\end{equation*}
for any $(t,x) \in [0,T] \times \mathbb{R}$. Hence, the drift, diffusion and jump components of the SDE (\ref{eq:dX-pi}) have at most linear growth. Furthermore, by (3) of Definition \ref{def:admissible-policy}, the drift, diffusion and jump components of the SDE (\ref{eq:dX-pi}) are also Lipschitz continuous. According to \cite{oksendal2005applied}, there exists a unique c\'adl\'ag adapted solution $X^\pi$ to the SDE (\ref{eq:dX-pi}) such that 
\begin{equation*}
    \mathbb{E}\left[\sup_{t\in[0,T]} (X_t^\pi)^2 \right] < \infty. 
\end{equation*}

\subsection{Proof of Theorem \ref{thm:verification}} \label{apdx:proof-of-verification-theorem}

First, by It\^o's formula, we have
\begin{equation}
    \mathbb{E}_{t,x} [g(T,X_T^{\pi^*})] = g(t,x) + \mathbb{E}_{t,x}\left[\int_t^T (\partial_t + \mathcal{L}^\pi) g(s, X_s^{\pi^*}) ds \right].
\end{equation}
Then, by (\ref{eq:HJB-g}) and (\ref{eq:HJB-terminal}), we get 
\begin{equation}
    g^{\pi^*}(t,x) \equiv g(t,x;\pi^*) := \mathbb{E}_{t,x} (X_T^{\pi^*}),
\end{equation}
which proves (\ref{eq:g-pi*}).

Next, we prove (\ref{eq:V=J-pi*}). Again by It\^o's formula, (\ref{eq:HJB-V}) and (\ref{eq:HJB-terminal}),
\begin{align*}
    V(t,x) = & \mathbb{E}_{t,x} \left[- \int_t^T (\partial_t + \mathcal{L}^\pi) V(s,X_s^{\pi^*})ds + V(T,X_T^{\pi^*}) \right] \\
    = & \mathbb{E}_{t,x} \Biggl[ 
    \int_t^T \lambda H(\pi_s^*) + \gamma g^{\pi^*}(s,X_s^{\pi^*}) \mathcal{L}^\pi g^{\pi^*}(s,X_s^{\pi^*}) - \frac{\gamma}{2} \mathcal{L}^\pi (g^{\pi^*}(s,X_s^{\pi^*}))^2 ds
    \Biggr] + \mathbb{E}_{t,x}[X_T^{\pi^*}]. 
\end{align*}

Note that for some $\pi \in \mathcal{A}, (t,x)\in[0,T]\times\mathbb{R}$, 
\begin{equation*}
    \frac{\gamma}{2} \partial_t(g^\pi(t,x))^2 = \gamma g^\pi(t,x) \partial_t g^\pi(t,x) = - \gamma g^\pi(t,x) \mathcal{L}^\pi g^\pi(t,x).
\end{equation*}

So, 
\begin{align*}
    V(t,x) = & \mathbb{E}_{t,x} \Biggl[ 
    \int_t^T \lambda H(\pi_s^*) ds - \frac{\gamma}{2} \int_t^T (\partial_t + \mathcal{L}^\pi) (g^{\pi^*}(s,X_s^{\pi^*}))^2 ds
    \Biggr] + \mathbb{E}_{t,x}[X_T^{\pi^*}] \\
    = & \mathbb{E}_{t,x} \Biggl[ 
    \int_t^T \lambda H(\pi_s^*) ds - \frac{\gamma}{2} (g^{\pi^*}(T, X_T^{\pi^*}))^2 +  \frac{\gamma}{2} (g^{\pi^*}(t,x))^2 
    \Biggr] + \mathbb{E}_{t,x}[X_T^{\pi^*}] \\
    = & \mathbb{E}_{t,x} \Biggl[
     \int_t^T \lambda H(\pi_s^*) ds - \frac{\gamma}{2} (X_T^{\pi^*})^2 + X_T^{\pi^*}
    \Biggr] + \frac{\gamma}{2} \left(\mathbb{E}_{t,x}(X_T^{\pi^*})\right)^2 \\
    = & J(t,x;\pi^*), 
\end{align*}
where the last equation is by (\ref{eq:J-pi}). 

Lastly, we show that $\pi^*$, the maximizer of (\ref{eq:HJB-V}), is an equilibrium policy. Consider a perturbed policy as in Definition \ref{def:equilibrium-policy}. That is, for a small time interval $\epsilon> 0$, $\pi^\epsilon = \{\pi^\epsilon_{s,y}(\cdot)\}_{(s,y)\in[0,T]\times\mathbb{R}}$, with
\begin{equation*}
    \pi _{s,y}^{\varepsilon }(\cdot),=%
    \begin{cases}
    \upsilon (\cdot), & (s,y)\in [t,t+\varepsilon) \times \mathbb{R}, \\ 
    \pi _{s,y}^{\ast }(\cdot), & (s,y)\in [t+\varepsilon ,T]\times \mathbb{R},
    \end{cases}%
\end{equation*}%
where $\nu(\cdot) \in \mathcal{P}(\mathbb{R})$. Under $\pi^\epsilon$, 
\begin{align*}
    J(t,x;\pi^\epsilon) = & \mathbb{E}_{t,x} \left[\int_t^{t+\epsilon} \lambda H(\nu) ds + \int_{t+\epsilon}^T \lambda H(\pi_s)ds + X_T^{\pi^\epsilon} - \frac{\gamma}{2} (X_T^{\pi^\epsilon})^2 \right] + \frac{\gamma}{2} \left(\mathbb{E}_{t,x}(X_T^{\pi^\epsilon}) \right)^2 \\
    = & \epsilon \lambda H(\nu) + \mathbb{E}_{t,x}(J(t+\epsilon, X_{t+\epsilon}^{\pi^\epsilon} ; \pi^\epsilon)) + \frac{\gamma}{2} (\mathbb{E}_{t,x}(X_T^{\pi^\epsilon}))^2 - \mathbb{E}_{t,x} \left[\frac{\gamma}{2} \left(\mathbb{E}_{t+\epsilon, X_{t+\epsilon}^{\pi^\epsilon}}(X_T^{\pi^\epsilon}) \right)^2 \right] \\
    = & \epsilon \lambda H(\nu) + \mathbb{E}_{t,x}(V(t+\epsilon, X_{t+\epsilon}^{\pi^\epsilon})) - \frac{\gamma}{2} \mathbb{E}_{t,x} \left[(g^{\pi^\epsilon}(t+\epsilon, X_{t+\epsilon}^{\pi^\epsilon}))^2 - (g^{\pi^\epsilon}(t,x))^2 \right]. 
\end{align*}
By It\^o's formula, we have 
\begin{align*}
    \mathbb{E}_{t,x}(V(t+\epsilon, X_{t+\epsilon}^{\pi^\epsilon})) - V(t,x) & = \mathbb{E}_{t,x} \left[\int_t^{t+\epsilon} (\partial_t +\mathcal{L}^\nu) V(s,X_s^\nu) ds \right]   \\
    & = \epsilon (\partial_t +\mathcal{L}^\nu) V(t,x) + O(\epsilon), 
\end{align*}
and 
\begin{align*}
    \mathbb{E}_{t,x} \left[\left(g^{\pi^\epsilon}(t+\epsilon, X_{t+\epsilon}^{\pi^\epsilon})\right)^2\right] - \left(g^{\pi^\epsilon}(t,x)\right)^2 & = \mathbb{E}_{t,x} \left[\int_t^{t+\epsilon} (\partial_t +\mathcal{L}^\nu) (g^{\nu}(s,X_s^\nu))^2 ds \right] \\
    & = \epsilon(\partial_t +\mathcal{L}^\nu) (g^{\nu}(t,x))^2 + O(\epsilon) \\
    & = \epsilon(-2 g^\nu(t,x) \mathcal{L^\nu} g^\nu(t,x) + \mathcal{L}^\nu (g^\nu(t,x))^2) + O(\epsilon). 
\end{align*}
Combining these together, we have
\begin{align*}
    & J(t,x;\pi^\epsilon) - V(t,x) \\
    & = J(t,x;\pi^\epsilon) - \mathbb{E}_{t,x}(V(t+\epsilon, X_{t+\epsilon}^{\pi^\epsilon})) + \mathbb{E}_{t,x}(V(t+\epsilon, X_{t+\epsilon}^{\pi^\epsilon})) - V(t,x) \\
    & = \epsilon \left(
    \lambda H(\nu) - \frac{\gamma}{2} \mathcal{L}^\nu (g^\nu(t,x))^2 + \gamma g^\nu(t,x) \mathcal{L}^\nu g^\nu(t,x) + (\partial_t + \mathcal{L}^\nu) V(t,x)
    \right) + O(\epsilon).
\end{align*}
We conclude from (\ref{eq:HJB-V}) that
\begin{equation*}
    \lim\sup_{\epsilon\downarrow 0} \frac{J(t,x;\pi^\epsilon) - J(t,x;\pi^*)}{\epsilon} = \frac{J(t,x;\pi^\epsilon) - V(t,x)}{\epsilon} \leq 0, 
\end{equation*}
which verifies that $\pi^*$ is an equilibrium policy.

\subsection{Proof of Theorem \ref{thm:equilibrium-policy}} \label{apdx:proof-of-equilibrium-policy-theorem}
According to the objective function (\ref{eq:J-pi}) and the definition of the auxiliary value function (\ref{eq:g-pi*}), we conjecture that the value function is quadratic in $x$ and the auxiliary value function is linear in $x$, i.e., 
\begin{align}
    & V(t,x) = A(t) x^2 + B(t) x + C(t), \\
    & g^\pi(t,x) = g(t,x;\pi) = x + h^\pi(t),
\end{align}
with $A(T) = C(T) = h^\pi(T) = 0$ and $B(T)=1$. 
Under this conjecture and by the infinitesimal generator (\ref{eq:generator}), for $(t,x)\in[0,T]\times\mathbb{R}$,
\begin{equation*}
    \mathcal{L}^\pi g^\pi(t,x) = \mathcal{L}^\pi (x+h^\pi(t)) = \int_{\mathbb{R}} u(\mu-r) \pi(u) du,
\end{equation*}
and
\begin{equation*}
\begin{split}
    \mathcal{L}^\pi (g^\pi(t,x))^2 & = \mathcal{L}^\pi (x^2 + (h^\pi(t))^2 + 2xh^\pi(t))  \\
    & = \int_{\mathbb{R}} u(\mu-r) \pi(u) du \cdot (2x + 2h^\pi(t)) + \int_{\mathbb{R}} \sigma^2 u^2 \pi(u)du \\
    & + \int_{\mathbb{R}} \int_{\mathbb{R}} [(g^\pi)^2(t, x+u(e^z-1)) - (g^\pi)^2(t, x) - u(e^z-1) \partial_x (g^\pi)^2] \pi(u) du \nu(dz) \\
    & = \int_{\mathbb{R}} u(\mu-r) \pi(u) du \cdot (2x + 2h^\pi(t)) + \int_{\mathbb{R}} \sigma^2 u^2 \pi(u)du + \int_{\mathbb{R}} u^2 \delta^2 \pi(u) du,
\end{split}
\end{equation*}
where $\delta^2 := \int_\mathbb{R} (e^z-1)^2 \nu(dz)$. The last equation is given by the fact that, for some function $\psi(t,x) \in \mathbb{C}^{1,2}([0,T]\times \mathbb{R})$ and $\psi$ is quadratic in $x$, the second order Taylor expansion yields $\psi(t, x+\Delta x) = \psi(t, x) + \psi_x(t, x) \Delta x + \frac{1}{2} \psi_{xx}(t,x) (\Delta x)^2$, and so, 

\begin{align*}
    & \int_{\mathbb{R}} \int_{\mathbb{R}} [\psi(t, x+u(e^z-1)) - \psi(t, x) - u(e^z-1) \psi_x] \pi(u) du\, \nu(dz) \\
    & = \int_{\mathbb{R}} \int_{\mathbb{R}} \frac{1}{2} u^2(e^z-1)^2 \pi(u) du\, \nu(dz) \cdot \psi_{xx}.
\end{align*}

So, from (\ref{eq:HJB-V}) and by (\ref{eq:generator}), 
\begin{align*}
    & (\partial_t + \mathcal{L}^\pi)V + \lambda H(\pi) + \gamma g^\pi \mathcal{L}^\pi g^\pi - \frac{\gamma}{2} \mathcal{L}^\pi (g^\pi)^2 \\
    = & V_t + \int_{\mathbb{R}} \left[u(\mu-r) V_x + \frac{1}{2} u^2\sigma^2 V_{xx} + \int_\mathbb{R}(V(t,x+u(e^z-1)) - V(t,x) - u(e^z-1)V_x) \nu(dz) \right] \pi_t(u) du \\
    & - \int_{\mathbb{R}} \lambda \pi_t(u) \log\pi_t(u) du 
    + \int_{\mathbb{R}} \gamma(x+h^\pi(t)) u(\mu-r) \pi(u) du \\
    & - \frac{\gamma}{2} \Biggl[\int_{\mathbb{R}} u (\mu-r) \pi_t(u) du \cdot (2x + 2h^\pi(t)) + \int_{\mathbb{R}} \sigma^2 u^2 \pi_t(u) du + \int_{\mathbb{R}} u^2 \delta^2 \pi_t(u) du \Biggr] \\
    = &  V_t + \int_{\mathbb{R}} \left[u(\mu-r) V_x + \frac{V_{xx} - \gamma}{2} u^2 (\sigma^2 + \delta^2)  -\lambda \log\pi_t(u)  \right] \pi_t(u) du. 
\end{align*}
The maximizer $\pi^*$ that attains the supremum of (\ref{eq:HJB-V}) is 
\begin{align}
    \pi^*_t(u) & = \exp\left( \frac{1}{\lambda}\left[u(\mu-r) V_x + \frac{V_{xx} - \gamma}{2} u^2 (\sigma^2 + \delta^2) - 1\right] \right) \notag \\
    & \propto \exp\left(\frac{1}{2}  \frac{(\sigma^2 + \delta^2) (V_{xx} - \gamma)}{\lambda}  \left[u^2 + \frac{2 (\mu-r) V_x}{(\sigma^2 + \delta^2) (V_{xx} - \gamma)}  u\right] \right) \notag \\
    & \sim N \left( -\frac{(\mu-r) V_x}{(\sigma^2 + \delta^2) (V_{xx} - \gamma)} ,  - \frac{\lambda}{(\sigma^2 + \delta^2) (V_{xx} - \gamma)}\right). \label{eq:pi*-with-V}
\end{align}
Substituting this back into (\ref{eq:HJB-V}), we have
\begin{align*}
    & V_t -\frac{(\mu-r)^2 V_x^2}{(\sigma^2 + \delta^2) (V_{xx} - \gamma)} + \frac{V_{xx} - \gamma}{2} (\sigma^2 + \delta^2) \left[- \frac{\lambda}{(\sigma^2 + \delta^2) (V_{xx} - \gamma)} + \frac{(\mu-r)^2 V_x^2}{(\sigma^2 + \delta^2)^2 (V_{xx} - \gamma)^2} \right] \\
    & +\frac{\lambda}{2} \log \left(- \frac{2\pi e \lambda}{(\sigma^2 + \delta^2) (V_{xx} - \gamma)} \right)  \\
    & = V_t -\frac{(\mu-r)^2 V_x^2}{2(\sigma^2 + \delta^2) (V_{xx} - \gamma)} + \frac{\lambda}{2} \log \left(- \frac{2\pi \lambda}{(\sigma^2 + \delta^2) (V_{xx} - \gamma)} \right) = 0.
\end{align*}
Under the conjectured forms of the value function and auxiliary value function, the left-hand-side of the above equation becomes 
\begin{align*}
    & A_t(t) x^2 + B_t(t) x + C_t(t) - \frac{(\mu-r)^2 (2A(t) x + B(t))^2}{2(\sigma^2 + \delta^2) (2 A(t) - \gamma)} + \frac{\lambda}{2} \log \left(-\frac{2\pi\lambda}{(\sigma^2+\delta^2)(2A(t) - \gamma)}\right) \\
    = & \Biggl\{ A_t(t) - \frac{4(\mu-r)^2 (A(t))^2}{2(\sigma^2+\delta^2)(2A(t) - \gamma)} \Biggr\} x^2
    +  \Biggl\{B_t(t) - \frac{4(\mu-r)^2 A(t) B(t)}{2(\sigma^2+\delta^2)(2A(t) - \gamma)} \Biggr\}x \\
    & + \Biggl\{C_t(t) - \frac{(\mu-r)^2  (B(t))^2}{2(\sigma^2+\delta^2)(2A(t) - \gamma)} + \frac{\lambda}{2} \log \left(-\frac{2\pi\lambda}{(\sigma^2+\delta^2)(2A(t) - \gamma)}\right) \Biggr\} = 0.
\end{align*}
So, $A, B, C$ can be solved by 3 ODEs.
\begin{align}
    & A_t(t) - \frac{4(\mu-r)^2 (A(t))^2}{2(\sigma^2+\delta^2)(2A(t) - \gamma)} = 0, \\
    & B_t(t) - \frac{4(\mu-r)^2 A(t) B(t)}{2(\sigma^2+\delta^2)(2A(t) - \gamma)} = 0, \\
    & C_t(t) - \frac{(\mu-r)^2  (B(t))^2}{2(\sigma^2+\delta^2)(2A(t) - \gamma)} + \frac{\lambda}{2} \log \left(-\frac{2\pi\lambda}{(\sigma^2+\delta^2)(2A(t) - \gamma)}\right) = 0.
\end{align}
Moreover, we know from (\ref{eq:HJB-g}) that
\begin{align*}
    (\partial_t + \mathcal{L}^\pi) g^\pi(t,x) 
    = & h^\pi_t(t) + \int_{\mathbb{R}} u (\mu-r) \pi_t(u) du \\
    = & h^\pi_t(t) - \frac{(\mu-r)^2 (2A(t)x + B(t))}{(\sigma^2 + \delta^2) (2A(t) - \gamma)} = 0.
\end{align*}
This tells us that $A(t) = 0, \forall t \in [0,T]$. So, 
\begin{equation}
    h^\pi_t(t) + \frac{(\mu-r)^2}{\gamma(\sigma^2 + \delta^2)} = 0, 
\end{equation}
which yields (\ref{eq:h-pi}). 
Then, it follows from the above ODEs that $B(t) = 1, \forall t \in [0,T]$ and 
\begin{equation}
    C_t(t) +  \frac{(\mu-r)^2}{2\gamma(\sigma^2+\delta^2)} + \frac{\lambda}{2} \log \left(\frac{2\pi\lambda}{\gamma(\sigma^2+\delta^2)}\right) = 0, 
\end{equation}
which yields (\ref{eq:C-pi}). 
This indicates that the value function is in fact linear in $x$ and is $V(t,x) = x + C(t)$. 
Substituting $V(t,x)$ into the equilibrium policy (\ref{eq:pi*-with-V}) yields
\begin{equation}
    \pi^*_t(u) \sim N\left(\frac{(\mu-r)}{\gamma (\sigma^2 + \delta^2)} ,  \frac{\lambda}{\gamma (\sigma^2 + \delta^2)} \right), 
\end{equation}
which is (\ref{eq:pi*}). Since $\pi^*_t(u)$ does not depend on the portfolio value $x$, it is easy to check that it is admissible according to Definition \ref{def:admissible-policy}, which completes the proof.

\subsection{Proof of Theorem \ref{thm:M-is-martingale}} \label{apdx:proof-M-is-martingale} 

Consider a parameterized policy $\pi^\theta$, its corresponding value function $V^\theta$ and auxiliary value function $g^\theta$, and a process $M^\theta$ defined in (\ref{eq:M-theta}). Then, by It\^o's formula, (\ref{eq:dX-theta}) and (\ref{eq:pi-theta}),  
\begin{align*}
    dM_t^\theta = & dX_t^\theta + C_t^\theta(t) dt - \gamma 
    g^\theta d (g^\theta) - \frac{\gamma}{2} 
    d[g^\theta] + \lambda H(\pi^\theta_t) dt \\
    = & \int_{\mathbb{R}} u (\mu_{true}-r) \pi^\theta_t(u) du dt + (...) dW_t + (...) d\Tilde{N}(dt,dz) \\
    & - \left(\frac{(\mu-r)^2}{2\gamma(\sigma^2+\delta^2)} + \frac{\lambda}{2} \log \left(\frac{2\pi \lambda}{\gamma(\sigma^2 + \delta^2)}\right) \right)dt 
    - \gamma 
    g^\theta [dX_t^\theta + h_t^\theta(t)dt] \\
    & - \frac{\gamma}{2} 
    d[X_t^\theta] + \frac{\lambda}{2} \log \left(\frac{2\pi e \lambda}{\gamma(\sigma^2 + \delta^2)}\right) dt \\
    %
    = & \int_{\mathbb{R}} u (\mu_{true}-r) \pi^\theta_t(u) du dt + (...) dW_t + (...) d\Tilde{N}(dt,dz) \\
    & - \left(\frac{(\mu-r)^2}{2\gamma(\sigma^2+\delta^2)} + \frac{\lambda}{2} \log \left(\frac{2\pi \lambda}{\gamma(\sigma^2 + \delta^2)}\right) \right)dt \\
    & - \gamma
    g^\theta \left[\int_{\mathbb{R}} (\mu_{true}-r) \pi^\theta_t(u) du dt - \frac{(\mu-r)^2}{\gamma(\sigma^2 + \delta^2)} dt + (...)dW_t + (...)d\Tilde{N}(dt,dz) \right] \\
    & - \frac{\gamma}{2} 
    \left[\int_{\mathbb{R}} \sigma_{true}^2 u^2 \pi_t(u) du dt + \int_{\mathbb{R}} \int_{\mathbb{R}} u^2 (e^z-1)^2 \pi_t(u) du dN(dt,dz) \right] \\
    & + \frac{\lambda}{2} \log \left(\frac{2\pi e \lambda}{\gamma(\sigma^2 + \delta^2)}\right) dt \\
    = & \int_{\mathbb{R}} u (\mu_{true}-r) \pi^\theta_t(u) du dt + (...) dW_t + (...) d\Tilde{N}(dt,dz) \\
    & - \left(\frac{(\mu-r)^2}{2\gamma(\sigma^2+\delta^2)} + \frac{\lambda}{2} \log \left(\frac{2\pi \lambda}{\gamma(\sigma^2 + \delta^2)}\right) \right)dt \\
    & - \gamma
    g^\theta \left[\int_{\mathbb{R}} (\mu_{true}-r) \pi^\theta_t(u) du dt - \frac{(\mu-r)^2}{\gamma(\sigma^2 + \delta^2)} dt + (...)dW_t + (...)d\Tilde{N}(dt,dz) \right] \\
    & - \frac{\gamma}{2} 
    \left[\int_{\mathbb{R}} \sigma_{true}^2 u^2 \pi_t(u) du dt + \int_{\mathbb{R}} \int_{\mathbb{R}} u^2 (e^z-1)^2 \pi_t(u) du \nu(dz)dt + (...)\Tilde{N}(dt,dz) \right] \\
    & + \frac{\lambda}{2} \log \left(\frac{2\pi e \lambda}{\gamma(\sigma^2 + \delta^2)}\right) dt \\
    = & \frac{(\mu_{true}-r)(\mu-r)}{\lambda(\sigma^2 + \delta^2)}dt + (...) dW_t + (...) d\Tilde{N}(dt,dz) \\
    & - \left(\frac{(\mu-r)^2}{2\gamma(\sigma^2+\delta^2)} + \frac{\lambda}{2} \log \left(\frac{2\pi \lambda}{\gamma(\sigma^2 + \delta^2)}\right) \right)dt \\
    & - \gamma 
    g^\theta 
    \left[\frac{(\mu_{true}-r)(\mu-r)}{\gamma(\sigma^2 + \delta^2)} - \frac{(\mu-r)^2}{\gamma(\sigma^2 + \delta^2)} \right]dt \\
    & - \frac{\gamma(\sigma_{true}^2+\delta_{true}^2)}{2} 
     \left(\frac{\lambda}{\gamma(\sigma^2 + \delta^2)} + \frac{(\mu-r)^2}{\gamma^2 (\sigma^2 + \delta^2)^2}\right) dt 
    + \frac{\lambda}{2} \log \left(\frac{2\pi e \lambda}{\gamma(\sigma^2 + \delta^2)}\right) dt.
\end{align*}
The drift term is zero when $\theta = \theta_{true}$, indicating that $M^\theta$ is a martingale.

\end{document}